\documentclass[a4paper,twoside]{article}

\usepackage{epsfig}
\usepackage{subcaption}
\usepackage{calc}
\usepackage{amssymb}
\usepackage{amstext}
\usepackage{amsmath}
\usepackage{amsthm}
\usepackage{multicol}
\usepackage{pslatex}
\usepackage{apalike}
\usepackage{ZhangCliffStyle}     

\begin{document}

\title{Market Impact in Trader-Agents: Adding Multi-Level Order-Flow Imbalance-Sensitivity to Automated Trading Systems}

\author{\authorname{Zhen Zhang\footnote{ORCID: 0000-0002-4618-6934} and Dave Cliff\footnote{ORCID: 0000-0003-3822-9364}}
\affiliation{Department of Computer Science, University of Bristol, Bristol BS8 1UB, U.K.}
\email{ad19014@bristol.ac.uk, csdtc@bristol.ac.uk}
}

\keywords{Market Impact, Adaptive Trader Agents, Financial Markets, Multi-Agent Systems.}

\abstract{Financial markets populated by human traders often exhibit so-called ``market impact", where the prices quoted by traders move in the direction of anticipated change, before any transaction has taken place, as an immediate reaction to the arrival of a large (i.e., ``block") buy or sell order in the market: traders in the market know that a block buy order is likely to push the price up, and that a block sell order is likely to push the price down, and so they immediately adjust their quote-prices accordingly. In most major financial markets nowadays very many of the participants are ``robot traders", autonomous adaptive software agents, rather than humans. This paper addresses the question of how to give such trader-agents a reliable anticipatory sensitivity to block orders, such that markets populated entirely by robot traders also show market-impact effects. This is desirable because impact-sensitive trader-agents will get a better price for their transactions when block orders arrive, and because such traders can also be used for more accurate simulation models of real-world financial markets. In a 2019 publication Church \& Cliff presented initial results from a simple deterministic robot trader, called ISHV, which was the first such trader-agent to exhibit this market impact effect. ISHV does this via monitoring a metric of imbalance between supply and demand in the market. The novel contributions of our paper are: (a) we critique the methods used by Church \& Cliff, revealing them to be weak, and argue that a more robust measure of imbalance is required; (b) we argue for the use of {\em multi-level order-flow imbalance} (MLOFI: Xu {\em et al.}, 2019) as a better basis for imbalance-sensitive robot trader-agents; and (c) we demonstrate the use of the more robust MLOFI measure in extending ISHV, and also the well-known AA and ZIP trading-agent algorithms (which have both been previously shown to consistently outperform human traders). Our results demonstrate that the new imbalance-sensitive trader-agents introduced in this paper do exhibit market impact effects, and hence are better-suited to operating in markets where impact is a factor of concern or interest, but do not suffer the weaknesses of the methods used by Church \& Cliff. We have made the source-code for our work reported here freely available on GitHub.
~\\
\vspace*{4em}
~\\
To be presented at the 13th International Conference on Agents and  Artificial Intelligence (ICAART2021), Vienna, 4th--6th February 2021.   
}

\onecolumn \maketitle \normalsize \setcounter{footnote}{0} \vfill

\section{\uppercase{Introduction}}
\label{sec:introduction}

\noindent Financial markets populated by human traders often exhibit so-called {\em market impact}, where the prices quoted by traders shift in the direction of anticipated change, as a reaction to the arrival of a large (i.e., ``block") buy or sell order for a particular asset: that is, mere knowledge of the presence of the block order is enough to trigger a change in the traders' quote-prices, before any transaction has actually taken place, because the traders know that a block buy order is likely to push the price of the asset up, and a block sell order is likely to push the price down, and so they adjust their quote-prices accordingly, in anticipation of the shift in price that they foresee coming as a consequence of the block-trade completing. This is bad news for the trader trying to buy or sell a block order: the moment she reveals her intention to buy a block, the market-price goes up; the moment she reveals her intention to sell, the price goes down. From the perspective of a block-trader, the market price {\em moves against her}, whether she is buying or selling, and this happens not because of the {\em price} she is quoting, but because of the {\em quantity} that she is attempting to transact. 

Block-traders' collective desire to avoid market impact has long driven the introduction of automated trading techniques such as ``VWAP engines" (which break block orders into a sequence of smaller sub-orders that are released into the market over a set period of time, with the intention of achieving a specific {\bf v}olume-{\bf w}eighted {\bf a}verage {\bf p}rice, hence VWAP), and has also driven the design of major new electronic exchanges such as London Stock Exchange's (LSE's) {\em Turquoise Plato} trading venue \cite{Turquoise2019}, in which block-traders are allowed to obscure the size of their blocks in a so-called {\em dark pool} market, with LSE's automated matching engine identifying one or more willing counterparties and only making full details of the block-trade known to all market participants after it has completed: see \cite{darkpoolStatistic2017} for further discussion.

Many of the world's major financial markets now have very high levels of automated trading: in such markets most of the participants, the traders, are ``robots" rather than humans: i.e., software systems for automated trading, empowered with the same legal sense of agency as a human trader, and hence ``software agents" in the most literal sense of that phrase. Given that these software agents typically replace more than one human trader, and given that those human traders were widely regarded to have required a high degree of intelligence (and remuneration) to work well in a financial market, it is clear that the issue of designing well-performing robot traders presents challenges for research in agents {\em and} artificial intelligence, and hence is a research topic that is central to the themes of the ICAART conference.

This paper addresses the question of best how to give robot traders an appropriate anticipatory sensitivity to large orders, such that markets populated entirely by robot traders also show market-impact effects. This is desirable because the impact-sensitive robot traders will get a better price for their transactions when block orders do arrive, and also because simulated market populated by impact-sensitive automated traders can be studied to explore the pros and cons of various impact-mitigation or avoidance techniques.  We show here that  well-known and long-established trader-agent strategies can be extended by giving them appropriately robust sensitivity to the imbalance between buy and sell orders issued by traders on the exchange, orders that are aggregated on the market's limit order-book (LOB), the data-structure at the heart of most electronic exchanges.  

To the best of our knowledge, the first paper to report on the use of an imbalance metric to give automated traders impact-sensitivity was the recent 2019 publication by Church \& Cliff, in which they demonstrated how a minimal nonadaptive trader-agent called {\em Shaver} (which, following the convention of practice in this field, is routinely referred to in abbreviated form via a psuedo-ticker-symbol: ``SHVR") could be extended to show impact effects by addition of an imbalance metric, and Church \& Cliff gave the name {\em ISHV} to their Imbalance-SHVR trader-agent \cite{ISHR2019}. SHVR is a trader-agent strategy built-in to the popular open-source financial exchange simulator called BSE \cite{BSEgithub}, which Church \& Cliff used as the platform for their research. Although Church \& Cliff deserve some credit for the proof-of-concept that ISHV provides, we argue here that the imbalance-metric they employed is too fragile for practical purposes because very minor changes in the supply and demand can cause their metric to swing wildly between the extremes of its range. One of the major contributions of our paper here is the demonstration that a much better, more robust, metric known as {\em multi-level order-flow imbalance} (MLOFI) can be used instead of the comparatively very fragile metric proposed by Church \& Cliff. Another major contribution of this paper is our demonstration of the addition of MLOFI-based impact-sensitivity to the very well-known and widely cited public-domain adaptive trader-agent strategies ZIP \cite{ZIP1997} and AA  \cite{AA2006,vytelingum2008}. Although our primary aim was to add impact sensitivity to these two machine-learning-based trader-agent strategies, we also demonstrate in this paper that ISHV can be altered/extended to use MLOFI, and our improvement of Church \& Cliff's work in that regard is an additional contribution of this paper. 

The extended versions of the AA, ZIP, and ISHV trader-agent strategies that we introduce here are named ZZIAA, ZZIZIP, and ZZISHV respectively. In this paper, after our criticism of Church \& Cliff's methods, we described more mathematically sophisticated approaches to measuring imbalance, which are more robust, and which we incorporate into our agent extensions. We then present results from testing our extended trader-agents on BSE, the same platform that was used in Church \& Cliff's work. Full details of the work reported here are available in \cite{zhenthesis}, and all the relevant source-code has been made freely available on {\em GitHub} \cite{zhengithub}.

The structure of the rest of this paper is as follows. Section 2 presents a broad overview of relevant background material: readers already familiar with automated trading systems and contemporary electronic financial exchanges can safely skip ahead straight to Section 3. In Section 3 we give a brief summary of Church \& Cliff's 2019 work and then provide our detailed critique of their core method, which we demonstrate to be significantly lacking in robustness, and we then describe our MLOFI approach in detail. Then Section 4 is where we describe the steps taken to add MLOFI-based impact-sensitivity to ZIP, AA, and ISHV; and the results from those extended trading algorithms are presented in Section 5. We discuss further work and draw our conclusions in Section 6.

\section{\uppercase{Background}}

Since the mid-1990s researchers in universities and in the research labs of major corporations such as IBM and Hewlett-Packard have published details of various strategies for autonomous trader-agents, often incorporating AI and/or machine learning (ML) methods so that the automated trader can adapt its behaviors to prevailing market conditions.  Notable trading strategies in this body of literature include: Kaplan's ``Sniper" strategy \cite{rust1992behaviour}; Gode \& Sunder's ZIC \cite{ZIC1993}; the ZIP strategy developed at Hewlett-Packard \cite{ZIP1997}; the GD strategy reported by Gjerstad \& Dickhaut \cite{GD1997} the MGD and GDX automated traders developed by IBM researchers \cite{TesauroDas2001,GDX2002}; Gjerstad's HBL \cite{Gjerstad2003}; Vytelingum's AA \cite{AA2006,vytelingum2008}); and the Roth-Erev approach (see e.g. \cite{Pentapalli2008}). However, for reasons discussed at length in a recent review of key papers in the field \cite{snashall_cliff_2019} this sequence of publications concentrated on the issue of developing trading strategies for orders that all had the same fixed-size quantity, and that quantity was always one. That is, {\em none} of the key papers listed here deal with trading strategies for outsize block orders, and none of them directly explore the issue of how an automated trader can best deal with, or avoid, market impact. 

Trader-agent strategies such as Sniper, ZIC, ZIP, GD and MGD were all developed to operate in electronic markets that were based on old-school open-outcry trading pits, as were common on major financial exchanges until face-to-face human-to-human bargaining was replaced by negotiation of trades via electronic communication media; but more recent work has concentrated on developing trading agents that issue bids and asks (i.e. quotations for orders to buy or to sell) to a centralised electronic exchange (such as a major stock-market like NYSE or NASDAQ or LSE) where the exchange's {\em matching engine} then either matches the trader's quote with a willing counterparty (in which case a transaction is recorded between the two counterparties, the buyer and the seller) or the quote is added to a data-structure called the {\em Limit Order Book} (LOB) that is maintained by the exchange and published to all traders whenever it changes. The LOB aggregates and anonymises all outstanding orders: it has two sides or halves: the bid-side and the ask-side. Each side of the book shows a summary of all outstanding orders, arranged from best to worst: this means that the bid-side is arranged in descending price-order, and the ask side is arranged in ascending price-order, such that at the ``top of the book" on the two sides the best bid and ask are visible. For all orders currently sat on the LOB, if there are multiple orders at the same price then the quantities of those orders are aggregated together, and often multiple orders at the same price will be later matched with a counterparty in a sequence given by the orders' arrival times, in a first-in-first-out fashion. The public LOB shows only, for each side of the book, the prices at which orders have been lodged with the exchange, and the total quantity available at each of those prices: if no orders are resting at the exchange for a particular price, then that price is usually omitted from the LOB rather than being shown with a corresponding quantity of zero. Illustrations of LOBs appear later in this paper, commencing with Figure~\ref{fig: mlofi_1}.

The difference between the price of the best bid on the LOB at time $t$ and the price of the best ask at $t$ is known as the {\em spread}. The mid-point of the spread (i.e. the arithmetic mean of the best bid and the best ask) is known as the {\em mid-price}  which is denoted here by $P_{\rm mid}$. The mid-price is very commonly used as a single-valued statistic to summarise the current state of the market, and as an estimate of what the next transaction price would be. However, the midprice pays no attention to the quantities that are bid and offered. If the current best bid is for a quantity of one at a price of \$10 and the current best ask is for a quantity of 200 at a price of \$20 then the mid-price is \$15 but that fails to capture that there is a much larger quantity being offered than being bid: basic microeconomics, the theory of supply and demand, would tell even the most casual observer that with such heavy selling pressure then actual transaction prices are likely to trend down -- in which case the mid-price of \$15 is likely to be an overestimate of the next transaction price. Similarly, if the bid and ask prices remain the same but the  {\em imbalance} between supply and demand is instead reversed, then the fact that there is a revealed desire for 200 units to be purchased but only one unit on sale at the current best ask would surely be a reasonable indication that transaction prices are likely to be pushed up by buying pressure, in which case the mid-price of \$15 will turn out to be an underestimate. This lack of quantity-sensitivity in the mid-price calculation leads many market practitioners to instead monitor the {\em micro-price}, denoted here by $P_{\rm micro}$, which is a quantity-weighted average of the best bid and best ask prices, and which does move in the direction indicated by imbalances between supply and demand at the top of the LOB: see, e.g., \cite{cartea_etal_2015}.

To the best of our knowledge the first impact-sensitive trading algorithm was {\em ISHV} \cite{ISHR2019}. ISHV is based on the {\em SHVR}\/ trader built into the popular {\em BSE}\/ public-domain financial-market simulator \cite{BSEgithub,cliff_bse_2018}. A SHVR trader simply posts the buy/sell order with its price set one penny higher/lower than the current best bid/ask. This single instruction gives it a parasitic nature, in the sense that it can mimic the price-convergence behaviour of other strategies being used by other traders in the market.  

Instead of shaving the best bid or offer by one penny, Church \& Cliff's ISHV trader instead chooses to shave by an amount $\Delta s$ which varies with $\Delta m$ defined in Equation~\ref{equ:ishv1}: 

\begin{equation}
\Delta m = P_{ \rm micro} - P_{\rm mid}
\label{equ:ishv1}
\end{equation}

The difference of the micro-price and the mid-price can identify the degree of supply/demand imbalance to a useful extent. If $\Delta m \approx 0$, there is no obvious imbalance in the market. If $\Delta m < 0$, then the quantities of the best bid and the best offer on the LOB indicate that supply exceeds demand and the subsequent transactions prices are likely to decrease; whereas $\Delta m > 0$ indicates that demand outweighs supply and subsequent transaction prices will have an upward tendency. 

\begin{figure}[htbp]
	\centering
	\includegraphics[width=0.25\textwidth]{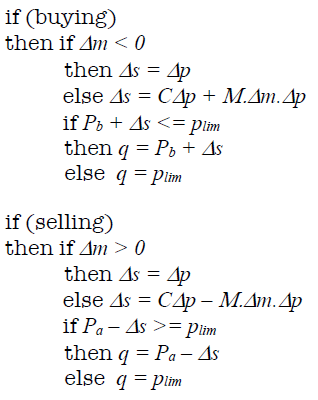}
	\caption{Pseudocode for the bidding behavior of ISHV, source from \cite{ISHR2019}.}
	\label{fig:ishvrule}
\end{figure}

The pseudocode for ISHV is shown in Figure~\ref{fig:ishvrule}. It implements a function that maps from $\Delta m$ to $\Delta s$ to determine how much it will shave off its price. For a buyer, if $\Delta m < 0$, it knows the price will shift in its favour and shaves its price as little as possible (the exchange's minimum tick-size  $\Delta p$ -- often one penny or one cent -- is chosen). However, if $\Delta m > 0$, ISHV ``believes" the later prices will be worse and attempts to shave a large amount off ($ C\Delta p+ M\Delta m\Delta p$). $C$ and $M$ are two constants that determine the SHVR's response to the imbalance (they are the $y-$intersect and gradient for a linear response function; nonlinear response functions could be used instead). Church \& Cliff showed that ISHV can identify and respond appropriately to the presence of a block order signal at the top of the LOB.

Church \& Cliff were careful to flag their ISHV trader as only a proof-of-concept (PoC): ISHV was developed to enable the study of coupled lit/dark trading polls such as LSE  {\em Turquoise Plato} system in commercial operation in London, as mentioned in the Introduction to this paper. Without impact-sensitive trader-agents, it is not possible to build agent-based models of contemporary real-world trading venues such as LSE Turquoise Plato. Having experimented further with Church \& Cliff's PoC system, we came to realise that there are severe limitations in ISHV as described in Figure~\ref{fig:ishvrule}: these limitations stem from the fact that Equation~\ref{equ:ishv1}, which is at the heart of ISHV, uses values {\em only found at the top of the LOB}. That is, Equation~\ref{equ:ishv1} involves only the price and quantity of the best bid and the best ask. As we will demonstrate in the next section, this makes the method introduced by Church \& Cliff so fragile that it is unlikely to be usable in anything but the simplest of simulation studies; as we show in the next section, for real-world markets it is necessary to look deeper into the LOB, to go beyond the top of the LOB.  

\section{\uppercase{Fragility of the LOB-top}}

For brevity, we will limit ourselves here to presenting a qualitative illustrative example which demonstrates how wildly fragile the Church \& Cliff method is. For a longer and more detailed discussion, see Chapter~3 of \cite{zhenthesis}.

Consider a situation in which the top of the LOB has a best bid price of \$10 and a best ask price of \$20, as before, and where the quantity at the best bid is 200 and at the best ask is 1. As we explained in the previous section, this huge imbalance between supply and demand at the top of the book indicate that the excess demand is likely to push transaction prices up in the immediate future. Church \& Cliff's ISHV does the right thing in this situation.  

Now consider what happens if the next order to arrive at the exchange is a bid for \$11 at a quantity of 1. Because this fresh bid is at a higher price than the current best bid, it is inserted at the top of the bid-side of the LOB. The previous best-bid, for 200 at \$10, gets shuffled down to the second layer of the LOB. At that point, the best bid and the best ask each show a quantity of one, and so ISHV acts as if there is no imbalance in the market, despite the fact when viewing {\em the whole LOB} it is clear that the quantity bid is now 201 (i.e. 1 at \$11 and 200 at \$10) while the ask quantity is still only 1: if anything, the imbalance has {\em increased} but ISHV reacts as if it had {\em disappeared} because ISHV looks only at the top of the LOB. 

There is more that could be said, but this should be enough to convince the reader that any impact-sensitive trader-agent algorithm that looks only at the data at the top of the LOB is surely going to get it wrong very often, because it is ignoring the supply and demand information, the quantities and the prices, which lie deeper in the LOB. What we introduce in the rest of this paper addresses this problem.

\section{\uppercase{Measuring Imbalance}}

A reliable metric is needed to capture the quantity imbalance between the supply side and the demand side, at multiple levels in the LOB (i.e., not just the top) and which can quantitatively indicate how much the imbalance will affect the market. We first discuss the Order-Flow Imbalance (OFI) metric introduced by \cite{cont2014price} and then describe the extension of this into a reliable Multi-Level OFI (MLOFI) metric very recently reported by \cite{xu2019multi}. After that, we show how MLOFI can be used to give robust impact-sensitivity to ISHV \cite{ISHR2019}, AA \cite{AA2006,vytelingum2008}, and ZIP \cite{ZIP1997}. AA and ZIP are of particular interest because in previous papers published at IJCAI and at ICAART it was demonstrated that these two trader-agent strategies can each reliably outperform human traders \cite{das_etal_ijcai,deluca_ijcai,deluca_icaart,deluca_szostek_cartlidge_cliff_2011}.

\subsection{Order Flow Imbalance (OFI)}
\label{sec: ofi1}

Cont et al.\ argued that previous studies modelling impact are extremely complex, and that instead a single factor, the order flow imbalance (OFI), can adequately explain the impact ($R^2 = 67\%$ in their research) \cite{cont2014price}. They indicated that $OFI$ has a positive linear relation with mid-price changes, and that the market depth $D$ is inversely proportional to the scope of the relationship. $OFI$ means the net order flow at the bid-side and the ask-side, and the market depth, $D$, represents the size at each bid/ask quote price.

To calculate the $OFI$ they focused on the ``Level 1 order book", i.e.\ the best bid and ask at the top of the LOB. Between any two events ($event_n$ and $event_{n-1}$), only one change happens in the LOB (check the condition from top to bottom, and from left to right; in other words, we should compare the change of price first and if the price does not change, then compare the change of quantity). Using $D\uparrow$ and $D\downarrow$ to respectively denote an increase and a reduction in demand; and $S\uparrow$ and $S\downarrow$ to denote an increase/decrease in supply, Cont et al.\ had:
\begin{eqnarray}
	p_{n}^{b}  > p_{n-1}^{b} & \vee & q^b_n > q^b_{n-1}  \implies D\uparrow \nonumber \\
	p_{n}^{b}  < p_{n-1}^{b} & \vee & q^b_n < q^b_{n-1}  \implies D\downarrow \nonumber  \\
	p_{n}^{a}  < p_{n-1}^{a} & \vee & q^a_n > q^a_{n-1}  \implies S\uparrow \nonumber \\
	p_{n}^{a}  > p_{n-1}^{a} & \vee & q^a_n < q^a_{n-1}  \implies S\downarrow \nonumber
\end{eqnarray}

Where $p^b$ is the best bid price; $q^b$  the size of the best bid price; $p^a$ the best ask price; and $q^a$ the size of the best ask price. The variable $e_n$ is defined to measure this tick change between two events, ($event_n$ and $event_{n-1}$), shown in Equation~\ref{equ: ofi1}, where $I$ can be regarded as a Boolean variable. 
	\begin{eqnarray}
	e_n & = & I_{\{p_{n}^{b} > p_{n-1}^{b}\}}q_n^b  - I_{\{p_{n}^{b} \leq p_{n-1}^b\}}q^b_{n-1}  \nonumber \\
	& & - I_{\{p_{n}^{a} < p_{n-1}^a\}}q^a_n  + I_{\{p_{n}^{a} \geq  p_{n-1}^{a}\}}q^a_{n-1}
	\label{equ: ofi1}
	\end{eqnarray}

The rules for $I$ are as follows, and only one of them will happen between any two consecutive events:
	\newline\newline
	1. if $p^b$ increases, $e_n$ = $q^b_n$\newline
	2. if $p^b$ decreases, $e_n$ = $-q^b_{n-1}$\newline
	3. if $p^a$ increases, $e_n$ = $q^a_{n-1}$\newline
	4. if $p^a$ decreases, $e_n$ = $-q^a_n$\newline
	5. if $p^b$ remains same and $q^b_n \neq q^b_{n-1}$, $e_n$ = $q^b_n-q^b_{n-1}$\newline
	6. if $p^a$ remains same and $q^a_n \neq q^a_{n-1}$, $e_n$ = $q^a_{n-1}-q^a_n$\newline\newline
	If $N(t_k)$ is the number of events during [$0$, $t_k$], then $OFI_k$ refers to the cumulative effect of $e_n$ that has occurred over the time interval [$t_k-1$, $t_k$], as shown in Equation~\ref{equ: ofi2}.
	\begin{equation}
	OFI_k = \sum_{n=N(t_{k-1})+1}^{N(t_k)}e_n
	\label{equ: ofi2}
	\end{equation}
	After this, a linear regression equation can be built, per Equation~\ref{equ: ofi3}, where $\Delta P_k = (P_k - P_{k-1})/ \delta $ and $\delta$ is the tick size (1 cent in Cont et al.'s experiments), $\beta$ is the price impact coefficient, and $\varepsilon_k$ is the noise term mainly caused by contributions from lower levels of the LOB:
	\begin{equation}
	\Delta P_k = \beta OFI_k + \varepsilon_k 
	\label{equ: ofi3}
	\end{equation}
	Moreover, Cont et al. stated that the market depth, $D$, is an important contributing factor to the fluctuations, and is inversely proportional to mid-price changes. They defined the average market depth, $AD_k$, in the ``Level 1 order book" as shown in Equation~\ref{equ: ofi4}; and $\beta$ can be measured by $AD_k$, shown in Equation~\ref{equ: ofi5}, where $\lambda$ and $c$ are constants and $v_k$ is a noise term.
	\begin{equation}
	AD_k = \frac{1}{2(N(T_k)-N(T_k)-1)}\sum_{N(T_{k-1})+1}^{N(T_k)}(q_n^B+q_n^A)
	\label{equ: ofi4}
	\end{equation}
	\begin{equation}
	\beta_k = \frac{c}{AD_k^{\lambda}}+v_k
	\label{equ: ofi5}
	\end{equation}
	
	Given equations \ref{equ: ofi3} and \ref{equ: ofi5}, the relationship between $\Delta P$ and $OFI$ and $AD$ is constructed as seen in Equation~\ref{equ: ofi6}, according to which, Cont et al.\ ran the linear regression by using the 21-trading-day data from 50 randomly chosen US stocks, and the average $R^2 = 67\%$. They demonstrated that $OFI$ is positive in relation to the change of mid-price. If $OFI$ $>$ 0, meaning a net inflow on the bid side or a net outflow on the ask side, the mid-price has a significantly increasing momentum, and the higher $OFI$ is, the more the mid-price will increase. Conversely, if $OFI$ $<$ 0, meaning a net outflow on the bid side or a net inflow on the ask side, the mid-price has a significantly decreasing momentum, and the lower $OFI$ is, the more the mid-price will decrease.
	\newline
	\begin{equation}
	\Delta P_k = \frac{c}{AD_k^{\lambda}} OFI_k + \epsilon_k
	\label{equ: ofi6}
	\end{equation}

OFI is clearly a useful metric, but it operates only on values found at the top of the LOB, i.e. the best bid and ask. In that sense, it is as sensitive to changes at the top of the book as is the Church \& Cliff $\Delta_{m}$ metric. Next we describe how OFI can be extended to be sensitive to values at multiple levels in the LOB, which gives us Multi-Level OFI, or MLOFI.

\subsection{Multi-Level Order Flow Imbalance}
Fortunately, \cite{xu2019multi} demonstrated how to measure multi-level order flow imbalance (MLOFI). A quantity vector, $v$, is used to record the OFI at each discrete level in the LOB: see Equation~\ref{equ: mlofi1}, where $m$ denotes the depth of price level in the LOB. The level-$m$ bid-price refers to the $m$-highest prices among bids in the LOB, and the level-$m$ ask-price refers to the $m$-lowest prices among asks in the LOB.
\newline
\begin{equation}
v=
\left(\begin{smallmatrix}
MLOFI_1\\ 
MLOFI_2\\ 
...\\
MLOFI_m\\
\end{smallmatrix}\right)
\label{equ: mlofi1}
\end{equation}

The time when an $n_{th}$ event occurs is denoted by $\tau_n$; $p^m_b(\tau_n)$ signifies the level-$m$ bid-price; $p^m_a(\tau_n)$ denotes the level-$m$ ask-price; $q^m_b(\tau_n)$ refers to the total quantity at the level-$m$ bid-price, and $q^m_a(\tau_n)$ refers to the total quantity at the level-$m$ ask-price.

Similar to the OFI defined in Section~\ref{sec: ofi1}, the level-$m$ OFI between two consecutive events occurring at times $\tau_{s}$ and $\tau_{n}$ ($s=n-1$) can be calculated as follows:
\newline
\begin{equation}
\Delta W^m(\tau_n) = 
\left\{\begin{matrix}
q^m_b(\tau_n), & {\rm if} \; p^m_b(\tau_n) > p^m_b(\tau_{s})\\ 
q^m_b(\tau_n) - q^m_b(\tau_{s}), & {\rm if} \; p^m_b(\tau_n) = p^m_b(\tau_{s})\\ 
-q^m_b(\tau_{m}), & {\rm if} \;p^m_b(\tau_n) < p^m_b(\tau_{s})\\ 
\end{matrix}\right.
\label{equ: mlofi2}
\end{equation}

\noindent
and

\begin{equation}
\Delta V^m(\tau_n) = 
\left\{\begin{matrix}
-q^m_a(\tau_{m}), & {\rm if} \; p^m_a(\tau_n) > p^m_a(\tau_{s})\\ 
q^m_a(\tau_n) - q^m_a(\tau_{s}), &{\rm if} \;p^m_a(\tau_n) = p^m_a(\tau_{s})\\ 
q^m_a(\tau_{n}), &{\rm if} \;p^m_a(\tau_n) < p^m_a(\tau_{s})\\ 
\end{matrix}\right.
\label{equ: mlofi3}
\end{equation}
\newline
where $\Delta W^m(\tau_n)$ measures the order flow imbalance of the bid side in the level-$m$ and $\Delta V^m(\tau_n)$ measures the order flow imbalance of the ask side in the level-$m$.

From equations~\ref{equ: mlofi2} and \ref{equ: mlofi3}, we can get the MLOFI in the level-$m$ over the time interval [$t_k-1$,$t_k$]:
\begin{equation}
MLOFI^m_k = \sum_{\{n|t_{k-1}<\tau_n<t_k\}}e^m(\tau_n)
\label{equ: mlofi4}
\end{equation}
\newline
where
\newline
\begin{equation}
e^m(\tau_n) = \Delta W^m(\tau_n) - \Delta V^m(\tau_n)
\label{equ: mlofi5}
\end{equation}

\begin{figure}[htbp]
	\centering
	\includegraphics[width=0.4\textwidth]{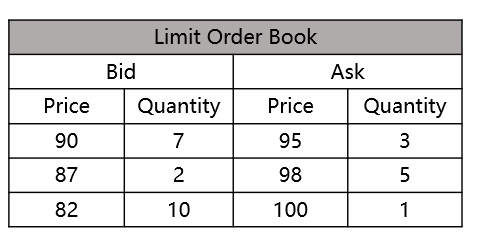}
	\caption{The LOB at time $t_{k-1}$.}
	\label{fig: mlofi_1}
\end{figure}

We now give four illustrative examples of the MLOFI mechanism in action. Figure~\ref{fig: mlofi_1} shows the situation of the LOB at time $t_{k-1}$, and there is only one event that occurs during the time interval [$t_{k-1}$,$t_{k}$], and here we'll only consider the 3-level OFI.

\subsubsection{Case 1: New order at level-1 of the LOB}

\begin{figure}[htbp]
	\centering
	\includegraphics[width=0.4\textwidth]{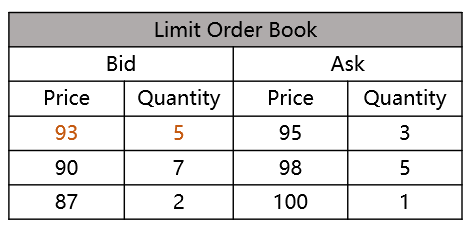}
	\caption{The LOB at time $t_{k}$: a new buy order comes.}
	\label{fig:mlofi_2}
\end{figure}

A new buy order comes into the LOB and occupies the best-bid position shown in Figure~\ref{fig:mlofi_2}.

\begin{itemize}
\item
Level-1: since $p^1_b(t_{k}) > p^1_b(t_{k-1})$ (i.e. $93 > 90$), $MLOFI^1_k = q^1_b(t_k) = 5$;
\item
Level-2: since $p^2_b(t_{k}) > p^2_b(t_{k-1})$ (i.e. $90 > 87$), $MLOFI^2_k = q^2_b(t_k) = 7$;
\item
Level-3: since $p^3_b(t_{k}) > p^3_b(t_{k-1})$ (i.e. $87 > 82$), $MLOFI^3_k = q^3_b(t_k) = 2$;
\end{itemize}

So, the quantity vector $v_k$ is:
\begin{equation}
v_k=
\left(\begin{smallmatrix}
5\\ 
7\\ 
2\\
\end{smallmatrix}\right)
\label{equ: mlofi6}
\end{equation}
 
All three numbers in $v_k$ are positive, which indicates the upward trend of the price.

\subsubsection{Case 2: Partial fulfillment or cancellation}

\begin{figure}[htbp]
	\centering
	\includegraphics[width=0.4\textwidth]{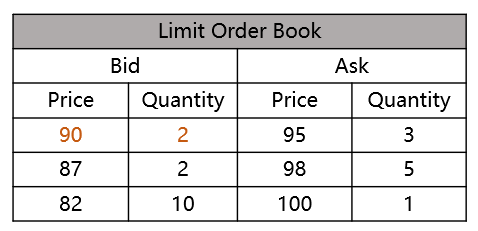}
	\caption{The LOB at time $t_{k}$: crossing the spread or a buy order cancellation.}
	\label{fig: mlofi_3}
\end{figure}

A new sell limit order crosses the spread, or a buy limit order at the best-bid position cancels. Figure~\ref{fig: mlofi_3} shows the resultant LOB.

For the level-1, as $p^1_b(t_{k}) = p^1_b(t_{k-1})$ (i.e. $90 = 90$), $MLOFI^1_k = q^1_b(t_k) - q^1_b(t_{k-1})$ $= 2-5= -3$;
\newline
For the level-2, as $p^2_b(t_{k}) = p^2_b(t_{k-1})$ (i.e. $87 = 87$), $MLOFI^2_k = q^2_b(t_k)- q^2_b(t_{k-1})$ $2-2 = 0 $;
\newline
For the level-3, as $p^3_b(t_{k}) = p^3_b(t_{k-1})$ (i.e. $82 = 82$), $MLOFI^2_k = q^3_b(t_k)- q^3_b(t_{k-1}) = 0$;
\newline\newline
So, the quantity vector $v_k$ is:
\begin{equation}
v_k=
\left(\begin{smallmatrix}
-3\\ 
0\\ 
0\\
\end{smallmatrix}\right)
\label{equ: mlofi7}
\end{equation}

Where $-3$ at Level 1 indicates a potential downward trend for the price, because the total demand on the bid side decreases.

\subsubsection{Case 3: full fulfillment or cancellation}

This is similar to Case 2, but (as illustrated in Figure~\ref{fig:mlofi_4}) assumes that all orders at Level 1 in the ask book  ($A_1$) are transacted by an incoming buy order, or that the order in  $A_1$ is cancelled. In this case, we need to consider the change on the ask side:

\begin{figure}[htbp]
	\centering
	\includegraphics[width=0.4\textwidth]{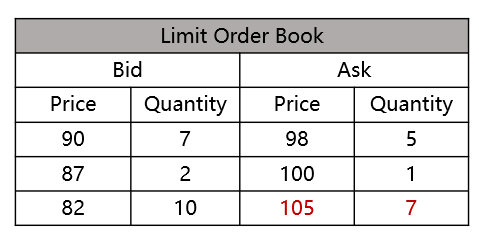}
	\caption{The LOB at time $t_{k}$: crossing the spread or a sell order cancellation.}
	\label{fig:mlofi_4}
\end{figure}

\begin{itemize}
\item
$A_1$:  $p^1_a(t_{k}) > p^1_a(t_{k-1})$ (i.e. $98 > 95$), $\implies \Delta V^1(t_k) = -q^1_a(t_{k-1}) = -3$; $MLOFI^1_k = -\Delta V^1(t_k) = 3$;
\item
$A_2$:  $p^2_a(t_{k}) > p^2_a(t_{k-1})$ (i.e. $100 > 98$), $\implies \Delta V^1(t_k) = -q^2_a(t_{k-1}) = -5$; $MLOFI^2_k = -\Delta V^1(t_k) = 5$;
\item
$A_3$:  $p^3_a(t_{k}) > p^3_a(t_{k-1})$ (i.e. $105 > 100$), $\implies \Delta V^1(t_k) = -q^2_b(t_{k-1}) = -1$; $MLOFI^3_k = -\Delta V^1(t_k) = 1$;
\end{itemize}
So, the quantity vector $v_k$ shown in Equation~\ref{equ: mlofi8} demonstrates that if the supply reduces or a buy has sufficient interest to transact, the price tends to go up.
 
\begin{equation}
v_k=
\left(\begin{smallmatrix}
3\\ 
5\\ 
1\\
\end{smallmatrix}\right)
\label{equ: mlofi8}
\end{equation}

\subsubsection{Case 4: New order at level-\textit{m} of the LOB}

\begin{figure}[htbp]
	\centering
	\includegraphics[width=0.4\textwidth]{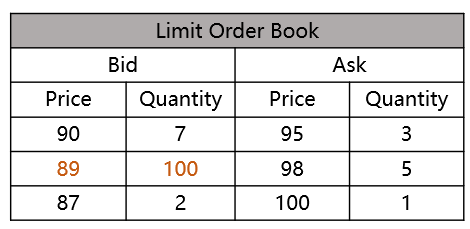}
	\caption{The LOB at time $t_{k}$: crossing the spread or a sell order cancellation.}
	~\label{fig: mlofi_5}
\end{figure}

Assuming now that a new large-sized order comes to the level-2 ask, if we only consider order flow imbalance in the top level of the LOB, we cannot detect this new block order. This is the reason why we choose to use MLOFI.
 
As there is no change in the level-1 bid, $MLOFI^1_k = 0$. Because a new order comes to the second-level bid, $p^2_b(t_{k}) > p^2_b(t_{k-1})$ (i.e. $89 > 87$) and $MLOFI^2_k = q^2_b(t_{k}) = 100$. Based on the same rule, $MLOFI^3_k = q^3_b(t_{k}) = 2$. So, the quantity vector $v_k$ is:
 
\begin{equation}
v_k=
\left(\begin{smallmatrix}
0\\ 
100\\ 
2\\
\end{smallmatrix}\right)
\label{equ: mlofi9}
\end{equation}

If we only care about first-level order flow imbalance, we get $OFI=0$. However, if we consider second and third levels, we get $MLOFI^2_k = 100$ and $MLOFI^3_k = 2$, which indicate a huge surplus on the demand side. If a trader can obtain this information and take action accordingly, it may result in larger profits or smaller losses.

\section{ZZIAA: AA traders with impact}

In this section we describe how ZZIAA is created, by the addition of MLOFI-style imbalance-sensitivity to the original AA trader strategy. Our intention for ZZIAA was to develop an ``impact-sensitive" module that is not deeply embedded into the original AA so that, if successful, this relatively independent module could also easily be applied to other trading algorithms. For this reason we chose the Widrow-Hoff delta rule to update the quote of the ZZIAA towards an impact-sensitive quote, as shown in Equation \ref{equ: iaa1}. The $p_{\rm AA}(t+1)$ is derived from the long-term and short-term factors using the information at time $t$ (see \cite{vytelingum2008}), and $\tau(t)$ is the target price computed with consideration of MLOFI:

\begin{equation}
	p_{\rm IAA}(t+1) = p_{\rm AA}(t+1)+ \Delta(t)
	\label{equ: iaa1}
\end{equation}
where
\begin{equation}
\Delta(t) = \beta (\tau(t)-p_{\rm AA}(t+1))
\label{equ: iaa2}
\end{equation}
and
\begin{equation}
\tau(t) = p_{\rm benchmark}(t) + o_{\rm offset}(t)
\label{equ: iaa0}
\end{equation}
 
The core of the IAA derivation is how to find $\tau(t)$, which consists of two parts, the benchmark price $p_{\rm benchmark}(t)$ and $o_{\rm offset}(t)$. The $p_{\rm benchmark}(t)$ depends on whether the mid-price exists. As Equation~$\ref{equ: iaa3}$ shows, if the mid-pice is available, we can set $p_{\rm benchmark}(t)$ as the mid-price, but if it is not, we set $p_{\rm benchmark}(t)$ as $p_{AA}(t+1)$, which can be obtained at time t.
\begin{equation}
p_{\rm benchmark}(t)=
\left\{\begin{matrix}
p_{\rm mid}(t),& {\rm if} \; \exists p_{\rm mid}\; \\ 
p_{\rm AA}(t+1),& {\rm if}\; \nexists p_{\rm mid}\; 
\end{matrix}\right.
\label{equ: iaa3}
\end{equation}
 
The $o_{\rm offset}(t)$ is derived from the MLOFI and the average depth. Assume that we consider $M$ numbers of levels MLOFI in the LOB, shown in Equation~\ref{equ: iaa4}, and that each MLOFI captures the last N events shown in Equation~\ref{equ: iaa5}. 
\begin{equation}
a(t)=
\left(\begin{smallmatrix}
{\rm MLOFI}_1(t)\\ 
{\rm MLOFI}_2(t)\\ 
...\\
...\\
{\rm MLOFI}_M(t)\\
\end{smallmatrix}\right)
\label{equ: iaa4}
\end{equation}
where
\begin{equation}
{\rm MLOFI}_M(t) = \sum_{n=1}^{N}e^m_n
\label{equ: iaa5}
\end{equation}
 
We can define the average market depth for $m$ levels in a similar way:

\begin{equation}
d(t)=
\left(\begin{smallmatrix}
{\rm AD}_1(t)\\ 
{\rm AD}_2(t)\\ 
...\\
...\\
{\rm AD}_M(t)\\
\end{smallmatrix}\right)
\label{equ: iaa6}
\end{equation}
where:
\begin{equation}
{\rm AD}_M(t) = \frac{1}{N}\sum_{n=1}^{N}\frac{q_{M_{n}}^a+q_{M_{n}}^b}{2}
\label{equ: iaa7}
\end{equation}
\newline
Knowing the quantity vector $a(t)$, we need a mechanism to switch this vector to a scalar. Similar to Equation~\ref{equ: ofi6}, we define the offset as Equation~\ref{equ: iaa8}.
\newline
\begin{equation}
v_{\rm offset} = \sum_{i=0}^{i=m-1}  \alpha^i\frac{c*{\rm MLOFI}_{(i+1)}(t)}{{\rm AD}_{(i+1)}(t)}
\label{equ: iaa8}
\end{equation}
\newline
where $\alpha$ is the decay factor (initialized as 0.8) and c is a constant (we use $c=5$). Note: if $AD_m(t)=0$, the item $\alpha^{m-1}\frac{c*{\rm MLOFI}_m(t)}{{\rm AD}_m(t)}$ will not be counted.
\newline\newline

To summarise, our work extends AA by the novel introduction of prior contributions to the econometrics of LOB imbalance from Cont et al.\ and of Xu et al.\ in the following ways:

\begin{itemize}

\item Cont et al.\ and Xu et al.\ run linear regressions to build their model and use statistical methods to test the significance of factors. The constants such as $c$ come from modelling real-world data. However our version does not run a linear regression and the constants such as $c$ and $\alpha$ are determined based on previous studies \cite{cont2014price,xu2019multi}. We can check the model's performance by exploring different values of constants.

\item In the prior work, $MLOFI_{M}(t)$ and $AD_{M}(t)$ are influenced by the events within a specified time interval. In contrast, in our work, $MLOFI_{M}(t)$ and $AD_{M}(t)$ are calculated based on the last $N$ events that occurred in the LOB, regardless of length of the time interval between successive events.

\end{itemize}

\section{\uppercase{Results}}

Because our MOLFI-based ``impact sensitive'' module added to AA was deliberately developed in a non-intrusive way, it can easily be replicated into any other algorithm. In this section we first show results from ZZIAA and then we follow those with results from adding the MLOFI module to ISHV (giving ZZISHV), and to ZIP (giving ZZIZIP). Because of space limitations, the performance comparisons shown here  focus on situations where the imbalance would cause a problem for the non-imbalance sensitive versions of the trader agents -- and we demonstrate that our extended trader agents are indeed superior. Extensive sets of further results are presented in \cite{zhenthesis}, which demonstrate that the extended trader-agents perform the same as the unextended versions in situations where there is no imbalance to be concerned about in the LOB.  

For each A:B comparison we ran 100 trials in BSE \cite{BSEgithub}, the same open-source simulator of a financial exchange that was used by Church \& Cliff. Each trial involved creating a market where there were $N$ traders of type A (e.g., ZIP) and $N$ traders of type B (e.g., ZZIZIP) who were allocated the role of buyers, and similarly $N$ of type A and $N$ of type B who were allocated the role of sellers. Thus one market trial involved a total of $4N$ trader-agents: for the results presented here we used $N=10$. As is entirely commonplace in all such experimental work, buyers were issued with assignments of cash, and sellers with assignments of items to sell, and each trader was given a private {\em limit price}: the price below which a seller could not sell and above which a buyer could not buy. The distribution of limit prices in the market determines that market's supply and demand curves, and the intersection of those two curves indicates the {\em competitive equilibrium price} that microeconomic theory tells us to expect transaction prices to converge to. 

Although very many of the previous trader-agent papers that we have cited here have monitored the {\em efficiency} of the traders' activity in the market, we instead monitored {\em profitability} (which only differs from efficiency by some constant coefficient). Each individual market trial would allow the traders to interact via the LOB-based exchange in BSE for a fixed period of time, and at the end of the session the average profit of the Type A traders would be recorded, along with the average profit of the Type B traders. In the results presented here we conducted 100 independent and identically distributed market trials for each A:B comparison, giving us 100 pairs of profitability figures. To summarise those results we plot as box-and-whisker charts the distribution of profitability values for traders of Type A, the distribution of profitability values for traders of Type B, and the distribution of profitability-difference values (i.e., for each of the 100 trials, for trial $t$ compute the difference between the profitability of Type A traders and the profitability of Type B traders in trial $i$). To determine whether the differences we observed were statistically significant, we used the Wilcoxon-Mann-Whitney U Test.

\subsection{ZZIAA}

\begin{figure}[htbp]
	\centering
	\includegraphics[width=0.45\textwidth]{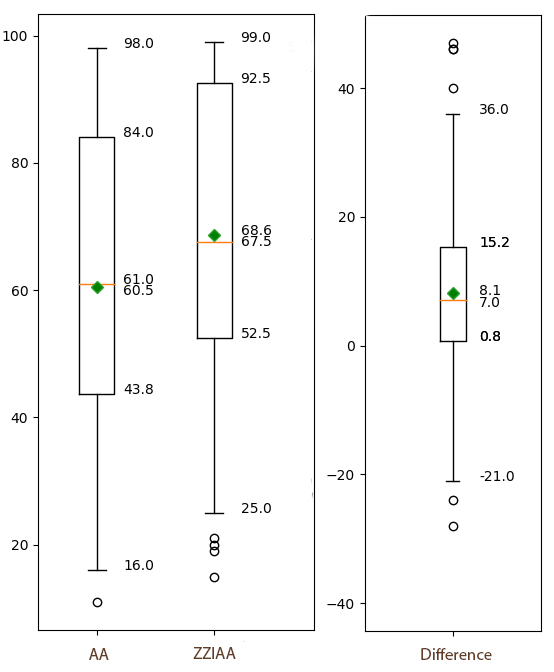}
	\caption{Profit distributions from original AA tested against ZZIAA.}
	\label{fig: box_newiaa2}
\end{figure}

Figure~\ref{fig: box_newiaa2} summarises the comparison data generated between AA and ZZIAA. In the U test, when comparing the ZZIAA with AA,  $p=0.007$ which meant that the profit difference between ZZIAA and AA was statistically significant. 
 
\subsection{Comparison of ZZISHV and ISHV}

\begin{figure}[htbp]
	\centering
	\includegraphics[width=0.6\textwidth]{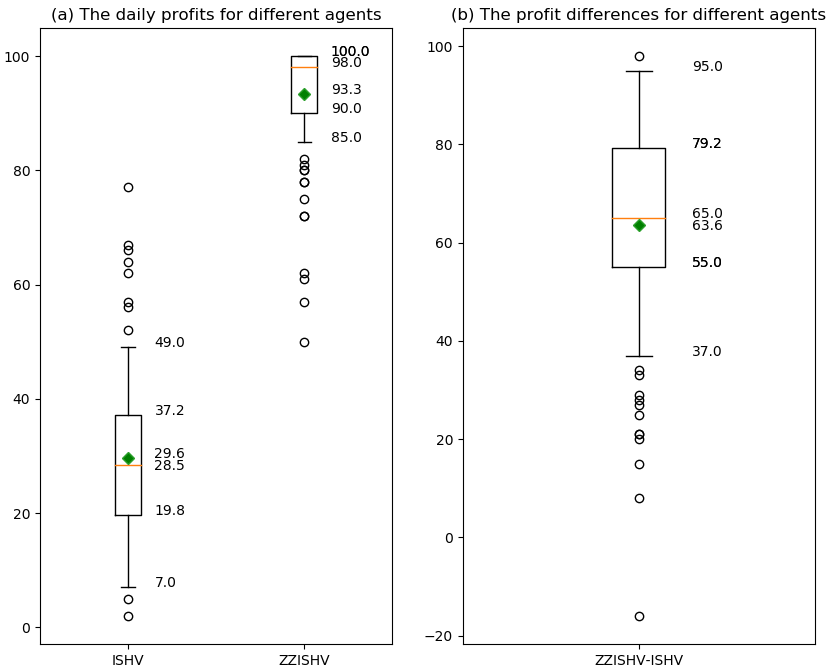}
	\caption{Performance of ZZISHV and ISHV when facing large-sized orders from the bid side.}
	\label{fig: box_zzishv1}
\end{figure}

We can see from Figure~\ref{fig: box_zzishv1} that the profit generated by ZZISHV was much greater than ISHV. However, this only means that ZZISHV is better than ISHV under this particular market condition, and this might not be the case under other market conditions. In the test, the outperformance of ZZISHV can easily be explained: as a seller, when ISHV met favourable imbalances, it worked like SHVR and posted a price one penny lower than the current best ask; in contrast, under the same condition, ZZISHV chose to set price $\Delta p$ higher than the current best ask and seek for transaction opportunities some time later. For example, assume that the current best ask is 70 and ISHV will post an order with the price equal to 69. Assume that ZZISHV gets the offset value equal to 20 from the ``impact-sensitive'' module, and the quoted price will be 90. 

The aim of both ISHV and ZZISHV is the same: to be sensitive to imbalances in the market. The former uses a function that maps from $\Delta_m$ to $\Delta_s$ to achieve this objective and $\Delta_m$ is generated based on the mid- and micro-prices in the market. In contrast, the latter uses MLOFI to achieve the goal. The biggest difference between ISHV and ZZISHV is that ISHV can only be sensitive to imbalances at the top of the LOB and the MLOFI mechanism helps ZZISHV to be sensitive to $m$-level imbalances on the LOB and thus detect them earlier than ISHV in some cases. The drawback comes in determination of appropriate parameter values for both ISHV and ZZISHV, where trial-and-error is the best current option. In the map function of ISHV ($\Delta s = C\Delta p \pm M.\Delta m.\Delta p$ if the imbalance is significant), the parameters $C$ and $M$ were somewhat arbitrarily set by \cite{ISHR2019} to C$=$2 and M$=$1. For ZZISHV, when quantifying MLOFI, we use Equation~\ref{equ: iaa8}, and the key parameter $c$ and decay factor $\alpha$ are artificially determined. We set $m=5$ (consistent with the result from \cite{cont2014price}) and $\alpha=0.8$. The optimal values of these parameters are not known; poor choices of these constants may cause agents to perform badly.

\subsection{Comparison of ZZIZIP and ZIP}
\begin{figure}[htbp]
	\centering
	\includegraphics[width=0.6\textwidth]{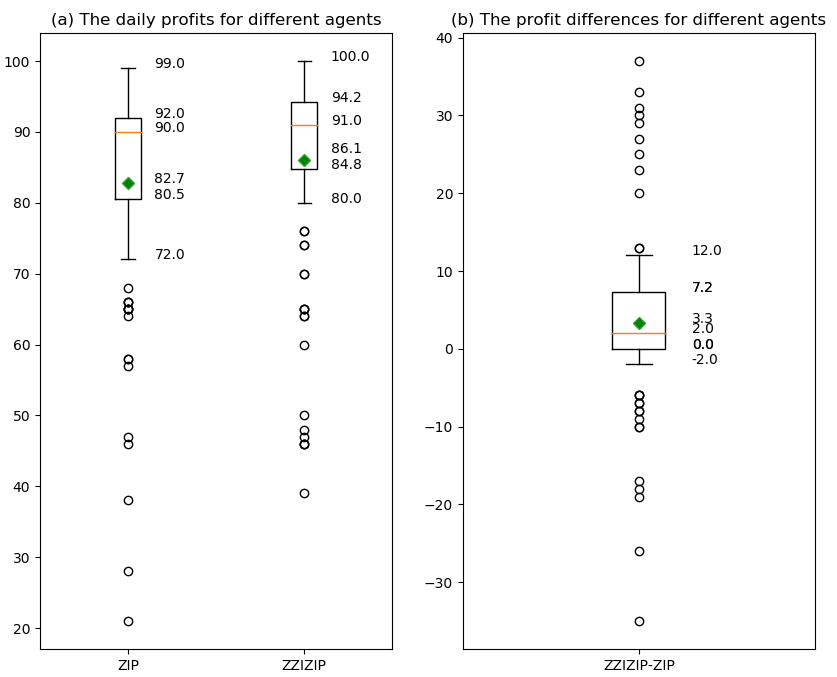}
	\caption{Performance of ZIP and ZZIZIP when facing large-sized orders from the bid side.}
	\label{fig: zzizip1}
\end{figure}

ZZIZIP is ZIP with the addition of the MLOFI module. In the example we present here, sellers will face an excess imbalance from the demand side.
The box plots in Figure~\ref{fig: zzizip1} illustrate the results: ZZIZIP has less variance than ZIP and their median profitability was slightly higher than that of ZIP; in the second figure, we can see that although there were some outliers on both the top and bottom, and the bottom whisker was located below zero, the whole box was distributed beyond zero. Employing the U Test, we got $p=0.002$ and can therefore conclude that the profit generated by ZZIZIP was statistically significantly greater than ZIP. Despite this, it is worth noting that the average difference in profitability is less than half of the difference between AA and IAA, given that other conditions remain unchanged. So, our next question is: what causes the smaller difference in profits between ZZIZIP and ZIP? 

To answer this, we need to examine how ZIP works. ZIP uses the Widrow-Hoff Delta rule to update its next quote-price towards its current target price. The current target price is based on the last quote price in the market. Due to this, the last quote price affects the bidding behaviour of ZIP considerably. In this test, on the ask side, the 10 ZIP sellers were not impact-sensitive and the 10 ZZIZIP sellers were. But, although the ZIP traders were not themselves impact-sensitive, they were affected by the quote prices coming from the ZZIZIP active in the same market, and so the ZIPs' quote prices approached the ZZIZIPs' to some extent. In other words, this adaptive mechanisms within  the non-impact-sensitive ZIP gave it a degree of impact-sensitivity, because it was influenced by the activities of the impact-sensitive traders in the market. In the test, if we treat ZZIZIP and ZIP as a group, the average profit generated is 84.82 (95\% CI: [82.16, 87.48]). If we replace 10 ZZIZIPs with 10 ZIPs (total 20 ZIP sellers), the average profit of ZIP is 79.21 (95\% CI: [77.11, 81.31]). With the presence of ZZIZIP, all sellers tend to make more profit.

\section{\uppercase{Discussion \& Conclusion}}

We know of no paper prior to \cite{ISHR2019} in which trader-agents are given a sensitivity to quantity imbalances between the bid and ask sides of the LOB. Such imbalances are often (but not always) caused by the arrival of one or more block orders on one side of the LOB. In this paper we have provided a constructive critique of Church \& Cliff's method, pointing out the extreme fragility of imbalance-sensitivity metrics like theirs that monitor only the top of the LOB. We then explained the OFI and MLOFI metrics of \cite{cont2014price} and \cite{xu2019multi} respectively, and demonstrated how MLOFI could be integrated within Vytelingum's AA trading-agent strategy to give ZZIAA. We demonstrated that ZZIAA performs extremely well: it performs the same as AA when there is no imbalance, and significantly outperforms AA in the presence of major LOB imbalance. We then showed how the imbalance-sensitivity mechanisms that we developed for ZZIAA can readily be incorporated into other trading-agent algorithms such as ZIP \cite{ZIP1997} and SHVR \cite{cliff_bse_2018}. Results from ZZIZIP and ZZISHV are similarly very good and further demonstrate that the mechanisms developed here have given robust imbalance-sensitivity to a range of trader-agent strategies. In future work we intend to explore the addition of  MLOFI-based impact-sensitivity to contemporary adaptive trader-agents based on deep learning neural networks \cite{lecalvezcliff2018,wraymeadescliff2020}. Complete details of the work described here are given in \cite{zhenthesis} and all of our relevant source-code for the system described here has been made freely available as open-source code on  {\em GitHub} \cite{zhengithub}, enabling other researchers to examine, replicate, and extend our work. 

\bibliographystyle{apalike}
{
\bibliography{ZhangCliffBib}}

\end{document}